\documentclass{aastex63}

\usepackage[figuresright]{rotating}
\usepackage{hhline}
\usepackage{multirow}
\submitjournal{The Astrophysical Journal}
\shorttitle{Recurrent Energetic Particle Enhancements}
\shortauthors{Krishnaprasad et al.}

\usepackage[T1]{fontenc}
\usepackage{makecell}

\begin{document}

\title{Recurrent Solar Energetic Particle Flux Enhancements Observed near Earth and Mars}

\author[0000-0002-3324-156X]{C. Krishnaprasad}
\affiliation{Space Physics Laboratory, Vikram Sarabhai Space Centre, Thiruvananthapuram 695022, 
India}
\affiliation{Cochin University of Science and Technology, Kochi 682022, India}

\author[0000-0002-0116-829X]{Smitha V. Thampi}
\affiliation{Space Physics Laboratory, Vikram Sarabhai Space Centre, Thiruvananthapuram 695022, 
India}

\author[0000-0003-1693-453X]{Anil Bhardwaj}
\affiliation{Physical Research Laboratory, Ahmedabad 380009, India}

\author[0000-0002-1604-3326]{Christina O. Lee}
\affiliation{Space Sciences Laboratory, University of California, Berkeley, CA 94720, USA}

\author{K. Kishore Kumar}
\affiliation{Space Physics Laboratory, Vikram Sarabhai Space Centre, Thiruvananthapuram 695022, 
India}

\author{Tarun K. Pant}
\affiliation{Space Physics Laboratory, Vikram Sarabhai Space Centre, Thiruvananthapuram 695022, 
India}

\correspondingauthor{C. Krishnaprasad, Smitha V. Thampi}
\email{kpchirakkil@gmail.com, smitha\_vt@vssc.gov.in}

\begin{abstract}

August 1 to November 15, 2016 period was characterized by the presence of Corotating Interaction 
Regions (CIRs) and a few weak Coronal Mass Ejections (CMEs) in the heliosphere. In this study we 
show recurrent energetic electron and proton enhancements observed near Earth (1 AU) and Mars 
(1.43--1.38 AU) during this period. The observations near Earth are using data from instruments 
aboard ACE, SOHO, and SDO whereas those near Mars are by the SEP, SWIA, and MAG instruments aboard 
MAVEN. During this period, the energetic electron fluxes observed near Earth and Mars showed prominent 
periodic enhancements over four solar rotations, with major periodicities of $\sim$27 days and 
$\sim$13 days.  Periodic radar blackout/weakening of radar signals at Mars are observed by 
MARSIS/MEX, associated with these  solar energetic electron enhancements. During this period, a weak 
CME and a High Speed Stream (HSS)-related interplanetary shock could interact with the CIR and 
enhance energetic proton fluxes near 1.43--1.38 AU, and as a result, $\sim$27 day periodicity in 
proton fluxes is significantly diminished at 1.43--1.38 AU. These events also cause unexpected 
impact on the Martian topside ionosphere, such as topside ionospheric depletion and compression 
observed by LPW and NGIMS onboard MAVEN. These observations are unique not only because of the 
recurring nature of electron enhancements seen at two vantage points, but also because they reveal 
unexpected impact of the weak CME and interplanetary shock on the Martian ionosphere, which provide 
new insight into the impact of CME-HSS interactions on Martian plasma environment.

\end{abstract}


\section*{Introduction} \label{sec:intro}

Corotating Interaction Regions (CIRs) are structures formed by the interaction between slow and fast 
solar wind streams having their primary origin in the solar coronal holes which persist for several 
solar rotations. The CIRs or Stream Interaction Regions (SIRs) are the most important cause of space 
weather disturbances during the declining and minimum phases of a solar cycle. The Coronal Mass 
Ejections (CMEs) are eruptions of plasma and magnetic field from the solar corona. In comparison to 
CIRs, they are short-lived transients, passing through the heliosphere with higher velocities. CMEs 
and CIRs are the major sources of energetic particles in the inner heliosphere. Recurrent 
CIR-accelerated ion events were previously observed by the Solar Electron and Proton Telescope 
onboard the twin STEREO spacecraft during the late phase of solar cycle 23 \citep{Herrero2009}. 
Interplanetary CMEs (ICMEs; interplanetary manifestation of CMEs typically remote-sensed by 
coronagraphs) are the major drivers of extreme space weather events on planets  and such events 
occur more frequently during solar maximum phase.  However, there are exceptions to this, like the 
large number of eruptions which occurred between 4 and 10 September 2017, including four eruptions 
associated with X-class flares in the decay phase of solar cycle 24, which has affected Earth 
\citep{Luhmann2018} and Mars \citep{Lee2018}. The declining phase of the weak solar cycle 24 is  
actually characterized by the presence of several slow CMEs, that is, CMEs with speeds below the 
typical solar wind speed of $\sim$400 km s$^{-1}$  (e.g. see the near-Earth ICME list, 
\url{http://www.srl.caltech.edu/ACE/ASC/DATA/level3/icmetable2.htm}, compiled by Richardson and Cane 
\citep{Richardson2010}), as well as by the occurrence of coronal holes, especially over low 
latitudes. Therefore, the interaction between slow/weak CMEs and High Speed Stream (HSS) is expected 
to happen during the late solar cycle 24.\\

The interaction of such slow/weak CMEs with other CMEs or the HSS can significantly change their 
properties while it propagates through the heliosphere \citep{Heinemann2019} and alter the 
geo-effectiveness \citep{Gopalswamy2009, He2018}. There are studies which shows that when two CMEs 
occur within $\sim$12 hours, the probability of having the second CME to be SEP-rich is quite high 
\citep{Gopalswamy2002, Gopalswamy2004}. It is also seen that the  CMEs having intense SEP 
enhancements are more likely to be preceded by another CME eruption \citep{Kahler2005}, though the 
exact process by which this interaction occur is still not very clear. When CIRs and ICMEs interact, 
they can also form Merged Interaction Regions (MIRs). Radial propagation and expansion of slow CMEs 
sometimes leads to formation of shocks \citep{Lugaz2017a}, also slow CMEs are accelerated during 
interaction with HSS in the heliosphere \citep{Gosling1996}. However, the impact of such CME-SIR 
interactions on the SEP characteristics as observed at different observer location is not well 
understood. Similarly, how the CME-HSS interactions may modify their impacts on Martian plasma 
environment is not reported so far.\\

Low solar activity affects the plasma system at Mars \citep{Ramstad2015, Hall2016, Sanchez2016}. 
During a solar minimum phase in March 2008, an ICME and SIR events impacted Mars and caused a strong 
compression of the magnetosheath and ionosphere \citep{Sanchez2017}. The dynamic response of Martian 
ionosphere to a CIR-related interplanetary shock (IPS) was studied by \citet{Harada2017}. They have 
found that subsequent to the sudden dynamic pressure enhancement, radar soundings recorded disturbed 
signatures in the topside ionosphere. In this study,  we present the observations of the periodic 
enhancements of solar energetic particles from vantage points near Earth (1 AU) and near Mars 
(during this period, Mars was between 1.43 and 1.38 AU, and the mean distance is 1.405 AU), for four 
consecutive solar rotations during 1 August \textendash 15 November 2016, along with observations of 
CME and HSS, and their impacts on the Martian plasma environment. This study is important because we 
do not have much information on how Mars is affected by space weather during the solar minimum, 
especially by CME-HSS interactions. 

\begin{figure}[ht] 
\centering
    \includegraphics[scale=1.45]{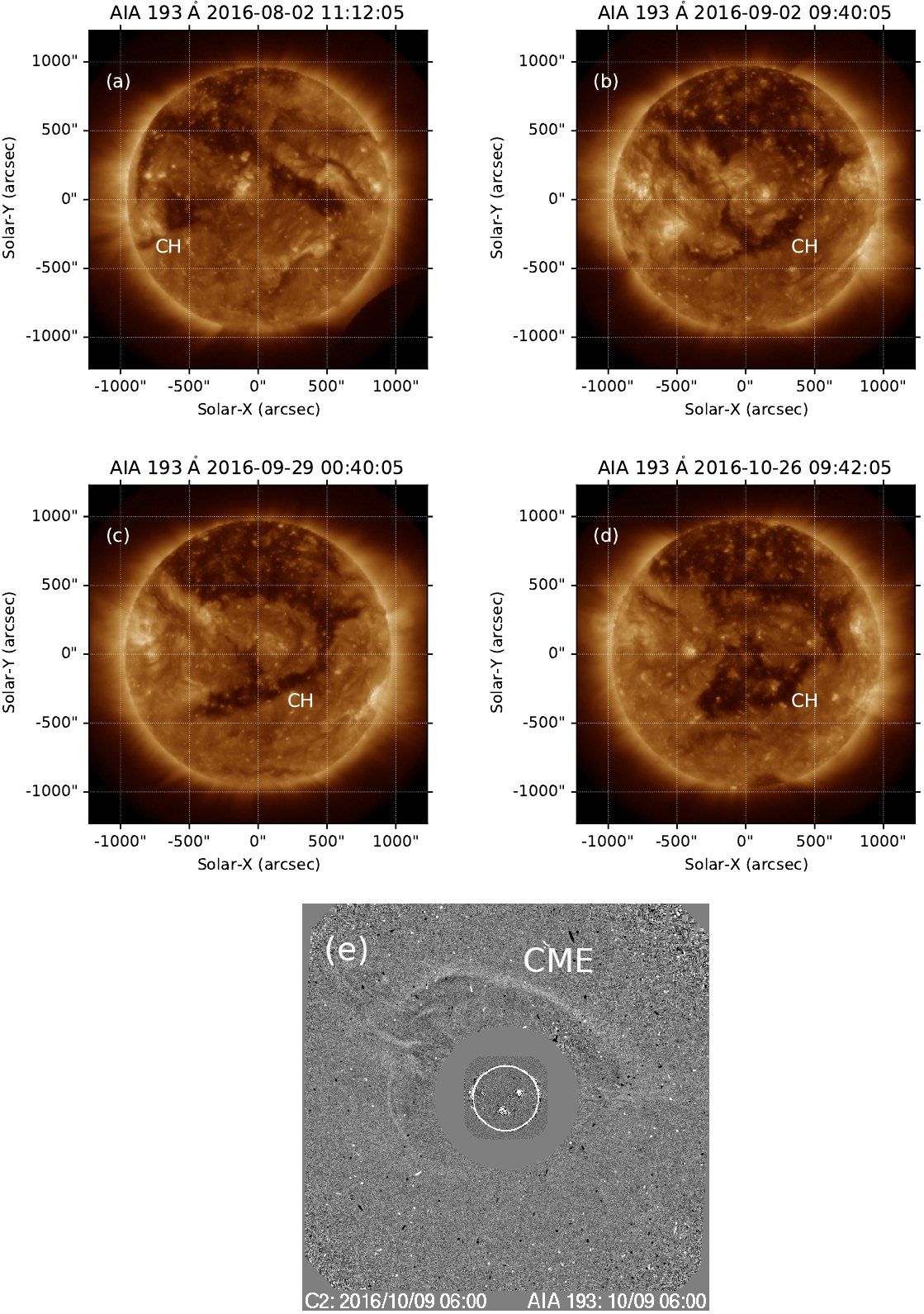}
    \caption{(a-d) The images of coronal hole on the solar disk by SDO/AIA for four successive solar 
rotations: (a) 2 August (DN 2), (b) 2 September (DN 33), (c) 29 September (DN 60), (d) 26 October 
(DN 87). (e) Difference image from SOHO LASCO C2 coronagraph superposed with Extreme ultraviolet 
Imaging Telescope image showing a weak CME eruption on 9 October (DN 70). DN: Day Number from August 
01, 2016.}
    \label{fig-allsolar}
\end{figure}

\begin{figure}[ht] 
    \centering
    \includegraphics[scale=0.65]{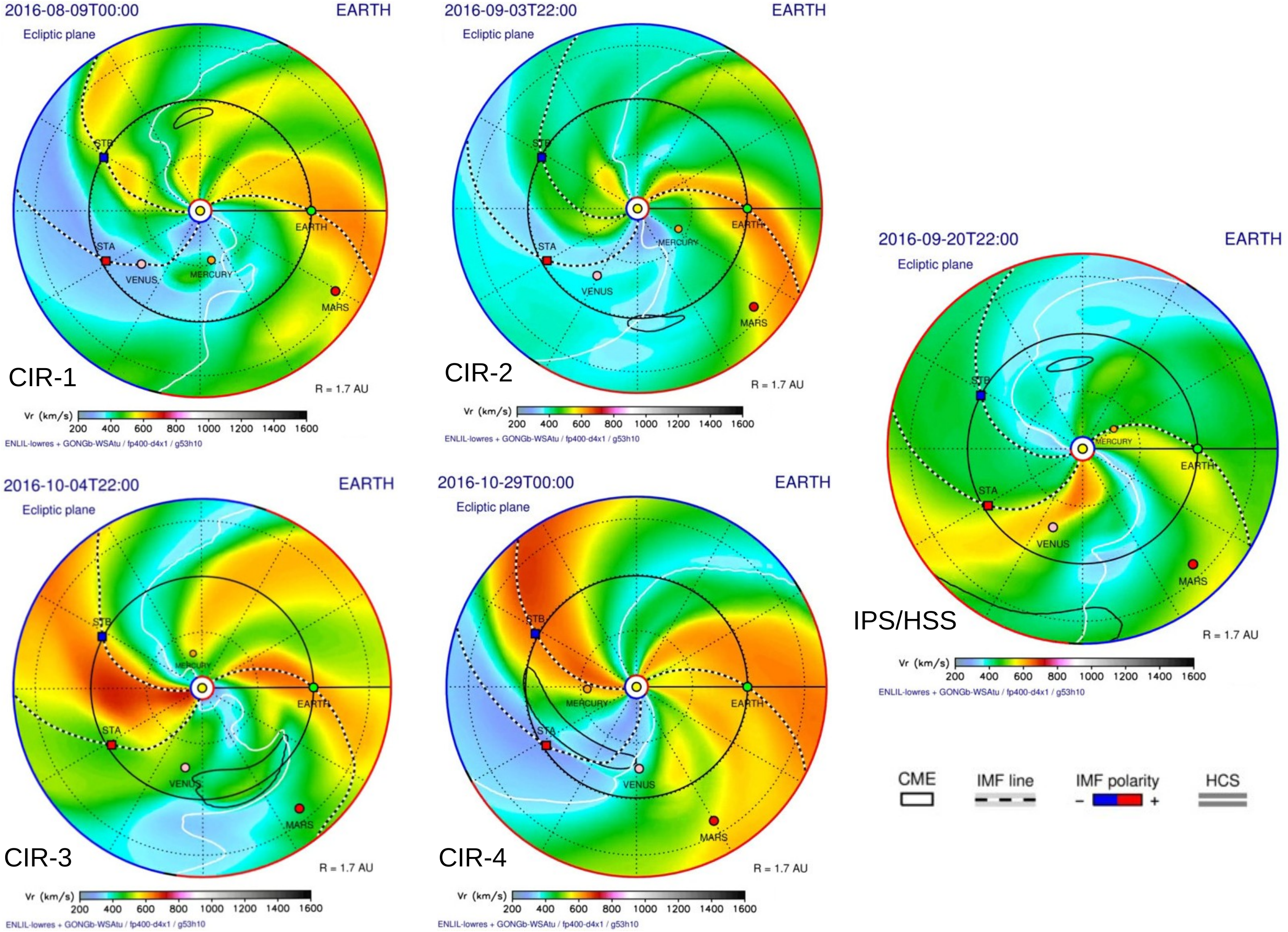}
    \caption{WSA-ENLIL simulation snapshots during CIR-1 (9 August, DN 9), CIR-2 (3 
September, DN 34), CIR-3 (4 October, DN 65), CIR-4 (29 October, DN 90), and IPS/HSS (20 September, DN 51) events at Earth. DN: Day Number from August 01, 2016.} \label{fig-enlil}
\end{figure}

\begin{figure}[ht] 
    \centering
    \includegraphics[scale=2.5]{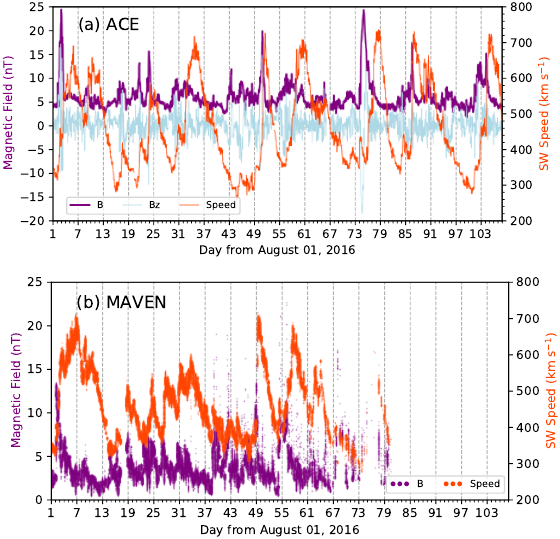}
    \caption{Temporal variation of the solar wind speed and Interplanetary Magnetic Field (IMF) 
observed by (a) ACE at L1 point and (b) MAVEN at 1.43--1.38 AU during 1 August -- 15 November 2016.} 
\label{fig-background}
\end{figure}

\begin{table}
  \centering
  \renewcommand{\arraystretch}{1.2}
  \begin{tabular}{|p{2cm}|c|c|c|c|c|c|}
    \hline
    \multirow{3}{2cm}{\textbf{Event}} & \multicolumn{3}{c|}{\textbf{Peak time at Earth} {[DN (month/day, hh:mm)]}} & \multicolumn{3}{c|}{\textbf{Peak time at Mars} {[DN (month/day, hh:mm)]}}\\
    \cline{2-7}
    & \textbf{Velocity} & \textbf{|B|} & \textbf{SEP} & \textbf{Velocity} & \textbf{|B|} & \textbf{SEP}\\
    \hline
\textbf{CIR-1} & 9 (08/09, 12:00) & 9 (08/09, 09:00) & 9 (08/09, 06:42) & 6 (08/06, 21:27) & 7 (08/07, 13:55) & 7 (08/07, 15:21)\\    
\hline
\textbf{CIR-2} & 34 (09/03, 16:04) & 36 (09/05, 16:04) & 36 (09/05, 08:52) & 33 (09/02, 17:45) & 34 (09/03, 20:52) & 35 (09/04, 20:38) \\    
\hline
\textbf{IPS/HSS} & 51 (09/20, 11:02) & 50 (09/19, 18:57) & 51 (09/20, 19:12) & 49 (09/18, 08:09) & 48 (09/17, 23:45) & 49 (09/18, 07:26) \\    
\hline
\textbf{CIR-3} & 65 (10/04, 11:00) & 65 (10/04, 12:00) & 63 (10/02, 02:52) & 62 (10/01, 16:33) & 61 (09/30, 11:45) & 61 (09/30, 11:45) \\    
\hline
\textbf{CME} & 74 (10/13, 01:55) & 74 (10/13, 22:04) & 73 (10/12, 03:07) & - & - & 76 (10/15, 07:12) \\    
\hline
\textbf{CIR-4} & 90 (10/29, 07:55) & 90 (10/29, 00:57) & 91 (10/30, 13:12) & - & - & 87 (10/26, 02:09) \\    
    \hline
  \end{tabular}
  \caption{Time (UTC) corresponding to maximum solar wind velocity, IMF |B|, and intensity of SEPs 
(SEP electrons for CIR/HSS while SEP protons for CME) observed at Earth and Mars. DN: Day Number 
from August 01, 2016.}
\end{table}

\section*{Data and Methods} \label{sec:Methods}
  
The solar observations are taken from Solar Dynamics Observatory (SDO)/Atmospheric Imaging Assembly 
(AIA;  \citet{Lemen2012}) (\url{https://sdo.gsfc.nasa.gov/}) and the Solar and Heliospheric 
Observatory (SOHO) Large Angle and Spectrometric Coronagraph (LASCO)-C2 (\url 
{https://cdaw.gsfc.nasa.gov/CME_list/}). These images show the presence of coronal holes as well as 
the CME eruptions. We have also used the Wang-Sheeley-Arge (WSA)\textendash ENLIL model 
\citep{Odstrcil2003, Mays2015} to numerically simulate the interplanetary solar wind plasma and 
magnetic field conditions, for a global heliospheric context for the solar events discussed, as well as for the relative planetary positions. The 
simulations used in the study are taken from ENLIL Solar Wind Prediction 
(\url{http://helioweather.net/}).  The simulations show the arrival of the high speed streams at 
both Earth and Mars, for four consecutive solar rotations and the arrival times are coinciding with 
the observed SEP electron enhancements.\\
  
The energetic particle measurements at the first Lagrangian point of the Sun-Earth system (L1) for the selected events are from the Electron Proton Alpha Monitor (EPAM) sensor onboard Advanced Composition Explorer (ACE) satellite \citep{Gold1998}. These  data are obtained from the ACE data center (\url{http://www.srl.caltech.edu/ACE/ASC/level2/}). The Deflected Electrons (DE) sensor  measures electron fluxes in 4 energy channels, that is, 0.038-0.053 MeV, 0.053-0.103 MeV, 0.103-0.175 MeV and 0.175-0.315 MeV.  The EPAM sensor also provides proton fluxes in the energy range from 47 keV to 4.75 MeV, in 8 channels and for the present work, we have used the ion fluxes measured in 4 channels in the energy ranges, 0.31-0.58 MeV, 0.58-1.05 MeV, 1.05-1.89 MeV, and 1.89-4.75 MeV by the Low Energy Magnetic Spectrometer (LEMS) of the EPAM sensor. The ACE Solar Wind Electron Proton Alpha Monitor (SWEPAM) and magnetic field experiment (MAG) measurements are used for interplanetary magnetic field (IMF) and solar wind speed observations at 1 AU.\\

The datasets from the Mars Atmosphere and Volatile EvolutioN (MAVEN) instruments are from the the Planetary Data System (\url{https://pds.nasa.gov/}). The solar wind speed and IMF values at 1.43--1.38 AU are obtained from the Solar Wind Ion Analyzer (SWIA; \citet{Halekas2015}) and Magnetometer (MAG; \citet{Connerney2015}) instruments aboard MAVEN spacecraft. SWIA is an energy and angular ion spectrometer that measures the energy and angular distributions of solar wind ions of energy between 25 eV and 25 keV with 48 logarithmically spaced energy steps. MAG is a fluxgate magnetometer that measures the intensity and direction of the IMF. The SWIA onboard moments and MAG measurements are used to compute the upstream solar wind and magnetic field parameters using the method by \citet{Halekas2017}. The method is based on the measured bulk flow speed |v|, proton scalar temperature T, altitude R, and normalized magnetic field fluctuation levels $\sigma_B /|B|$, where $\sigma_B$ is the root-sum-squared value of the 32 Hz fluctuation levels in all three magnetic field components over a 4 s interval.  To select undisturbed solar wind intervals, points with |v| $>$ 200 km/s, $\sigma_B /|B|$ $<$ 0.15, R $>$ 500 km, and $\sqrt{T}/|v|$ $<$ 0.012 are chosen \citep{Halekas2017}. In the later phase, that is, after $\sim$ October 20, 2016, the orbit of MAVEN was inside the Martian bow shock and hence it was not observing the `{pure}' solar wind condition, and therefore the solar wind speed and IMF could not be deduced. The level 2, version 01, revision 01 (V01\_R01) data of SWIA and level 2, version 01, revision 01/02 (V01\_R01/R02) data of MAG are used for analysis.\\ 

The SEP fluxes are obtained from the Solar Energetic Particle (SEP) instrument aboard MAVEN. This instrument consists of two identical sensors, SEP 1 and SEP 2, each consisting of a pair of double\textendash ended solid\textendash state telescopes  to measure 20 keV\textendash 200 keV electrons and 30 keV\textendash 6 MeV ions in four orthogonal view directions \citep{Larson2015}. The data used in this study  are the ion and electron data in the form of differential energy fluxes measured by the SEP 1 sensor in the 1F direction, that typically views the Parker spiral direction \citep{Larson2015}. Both SEP 1 and SEP 2 sensors in the forward and reverse facing FOVs (field of views) observed the recurrent SEP enhancements. The Level 2, version 04, revision 02  (V04\_R02) data of SEP are used. The Langmuir Probe and Waves (LPW; \citet{Andersson2015}) instrument onboard MAVEN is used for the in situ electron density measurements [Level 2, version 3, revision 01 (V03$\_$R01)]. The electron densities in the LP mode are derived based on I--V characteristics. The Neutral Gas and Ion Mass Spectrometer (NGIMS; \citet{Mahaffy2014}) observations of MAVEN are used to understand the variations of O$_2^+$ and O$^+$ ion densities in the Martian ionosphere. NGIMS is a quadrupole mass spectrometer which measures the composition of neutrals and thermal ions, in the mass range 2-150 amu with unit mass resolution. The NGIMS Level 2 ion data version 08, revision 01 (V08$\_$R01) are used.\\

The Mars Advanced Radar for Subsurface and Ionosphere Sounding (MARSIS) on board Mars Express (MEX) mission is a nadir-looking pulse-limited radar sounder operating at both active ionospheric sounding (AIS) as well as subsurface sounding modes \citep{Gurnett2008, Orosei2015}. MARSIS consists of a 40 m antenna, with associated radio transmitter and receiver. In AIS mode, the operation is that of a swept-frequency radar sounder, with 160 frequencies chosen from 0.1 to 5.5 MHz in roughly logarithmic spacing. The transmitter sends a 91 $\mu$ s tone at 127 pulses per second rate. The frequency sweep takes 7.3 s to complete the 0.1-5.5 MHz range \citep{Orosei2015}. Even in the AIS mode,  the transmitted frequencies can  be reflected from the Martian surface if they are larger than the critical plasma frequency of the Martian ionosphere. However this is true only for the terminator and nightside observations, while the signal absorption at lower solar zenith angles such as $<$70 deg. could be due to the absorption by the robust Martian dayside ionosphere. For the present study, we have used data from  the AIS mode, but the echoes corresponding to the 4 MHz frequency are actually due to the surface reflections. The MARSIS/MEX data are given in the PDS-3 format and are downloaded from the ESA-PSA archive (\url {https://www.cosmos.esa.int/web/psa/mars-express}).\\

\begin{sidewaysfigure}[ht] 
\hspace{-2cm}
\includegraphics[scale=2.2]{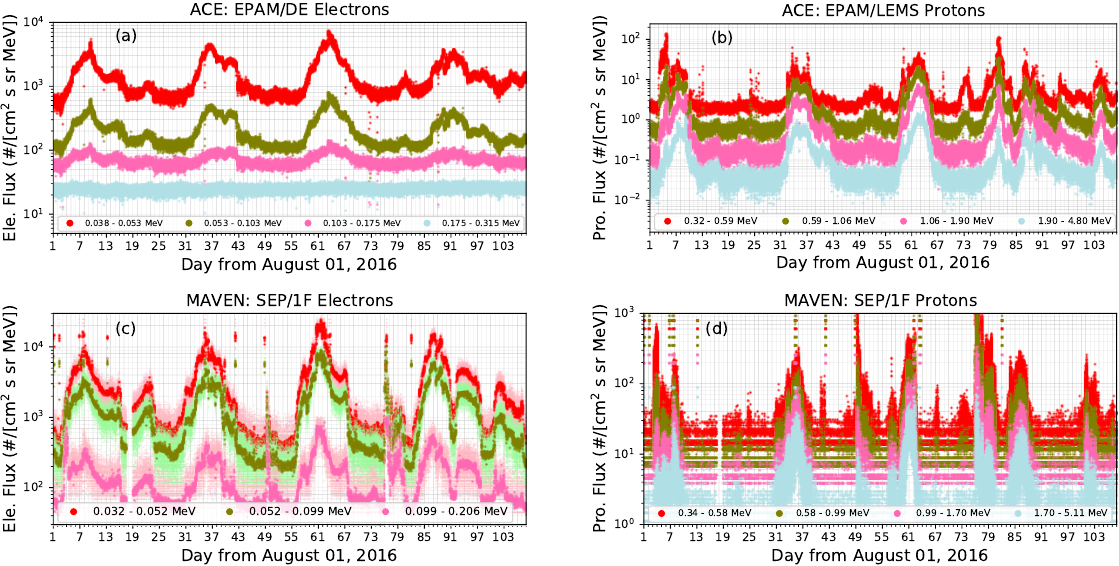} \caption{Temporal 
variation of the Solar Energetic Particle (SEP) fluxes during 1 August -- 15 November 2016. (a) 
Electron fluxes observed by ACE,  (b) Proton fluxes observed by ACE, (c) Electron fluxes  observed 
by MAVEN, (d) Proton fluxes observed by MAVEN.} \label{fig-ace-maven} 
\end{sidewaysfigure}

\section*{Overview of Events} \label{sec:SolarObs}

Figure 1(a-d) shows the presence of persistent coronal holes on the solar disk as seen in the 193 
\AA{} image from the AIA onboard SDO.  These are the sources of the continuous HSS resulting in 
corotating interaction region events. Due to the similar relative observer positions for Earth and 
Mars during this period, we can consider these as stationary heliospheric observers while the 
coronal hole source continuously emits high speed streams as it rotates in a right\textendash handed 
direction, and hits the observers with the $\sim$27\textendash day rotation period of the Sun. The 
arrival of these streams corresponds to enhancements in the SEP fluxes, observed at both L1 point 
and near Mars (see Table 1). The differences in the stream arrival times are consistent with the 
relative location of the planets in the heliosphere, and the heliolongitude of the source location.  
Figure 1e shows the snapshot of a CME eruption observed by  LASCO C2 coronagraph during 8\textendash 
9 October, 2016 (DN 69 to 70, here DN denotes the Day Number from August 01, 2016). There were no 
obvious signatures in the low corona, and therefore this is probably a `stealth CME' \citep{He2018}. 
Unlike a typical CME eruption that leaves one more signatures in the solar disk, a stealth CME will 
have no obvious low coronal signature. The AIA images suggest that there are coronal holes near the 
solar equator during this time \citep{He2018}.  The HSS from the coronal holes may influence the 
propagation of the CME in the heliosphere \citep{Liu2016}. However, the arrivals of the HSS and the 
CME eruption at both the planets are marked by distinct enhancements in SEP fluxes as described 
later. Also, there was an HSS-related interplanetary shock during 17 to 20 September (DN 48 to 51). 
The science events list in MAVEN science data center shows the arrival of interplanetary shock and 
low energy SEP ions accelerated by a strong solar wind 
stream during this period (\url{https://lasp.colorado.edu/maven/sdc/public/}). In addition to 
these, there were weak/moderate CME eruptions  during the beginning of August, towards the end of 
September and during early October, some of them being west\textendash limb eruptions, and these 
might have been additional sources of SEPs at Earth and Mars, through magnetic connections.\\ 

Figure 2 shows the WSA-ENLIL simulation snapshots during the passage of CIR (at Earth) for four consecutive solar rotations, as well as during the IPS/HSS event at Earth. It can be observed that Mars being behind the Earth, will observe the CIR/HSS first compared to Earth, owing to the counter-clockwise rotation of the Sun. The stealth CME event of 8\textendash 9 October is not captured in the general simulation. However, the CME event is observed by SOHO/LASCO-C2, which is shown in Figure 1(e). The time-dependent WSA-ENLIL+Cone global 3D MHD model with graduated cylindrical shell (GCS) results as input (including the time when the CME crosses the inner boundary at 21.5 R$_\odot$) do capture the presence of this stealth-type CME, as shown in Figure 6 of \citet{He2018}. Table 1 shows the times at maximum solar wind speed, IMF B$_{tot}$, and SEP intensities are observed at Earth and Mars during the prominent events in the period (for the case of CIR/HSS events, the time given is the time of maximum solar energetic electron intensities observed, while for the CME event, the time given is the time of maximum solar energetic proton intensities observed, because during the CME event, the proton enhancement is more prominent).\\

Figure 3a shows the variation of IMF and solar wind speed as observed at L1 point (near Earth) by 
ACE satellite, during 01 August \textendash 15 November 2016 (DN 1 to 107). The day numbers marked 
in x-axis starts from August 01, 2016 and the same convention is followed in the subsequent plots as 
well. Periodic enhancement in solar wind speed is seen as a result of the periodic arrivals of the 
high speed streams. The z\textendash component of the IMF as well as total B show largest 
fluctuations coincident with the arrival of the CIRs, followed by fluctuations of smaller magnitude, 
a typical feature of CIR  events near 1 AU \citep{Borovsky2006}. Here the B$_z$ is also shown to 
understand when the IMF turns southward, and that would cause a magnetic reconnection and a 
geo-effective storm. Figure 3b shows the variation of IMF and solar wind speed as observed by the 
MAG and SWIA instruments aboard MAVEN located near 1.43--1.38 AU. The total magnetic field increases 
 coincident to the arrival of the high speed stream. The solar wind speed shows periodic 
enhancements similar to that observed near Earth. In addition to the stream arrivals, there are 
clear signatures of the interplanetary shock and CME arrivals at both the locations during 17-18 
September (DN 48-49) and 14-15 October 2016 (DN 75-76) respectively. The event arrivals can be 
identified with the enhancements in solar wind speed and magnetic field (Figure 3b).\\ 

\begin{figure}[h] \centering \includegraphics[scale=2.4]{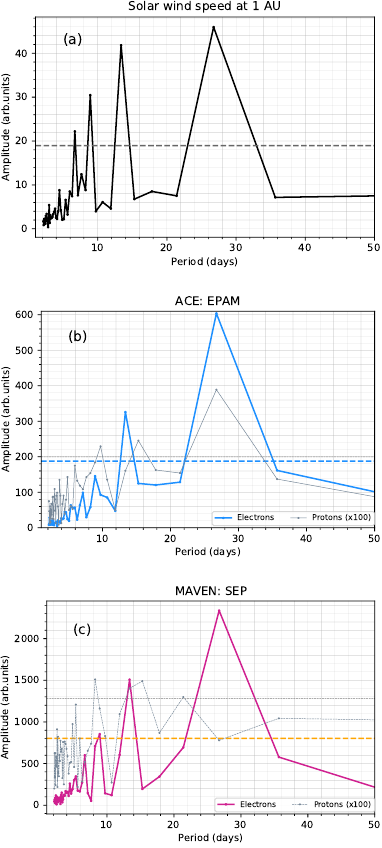} \caption{FFT spectra of 
(a) solar wind speed near 1 AU (ACE: SWEPAM observations) (b) FFT spectra of energetic particles 
near 1 AU (ACE: EPAM observations) where the electron spectra are shown in blue and the proton 
spectra are shown in gray. Note that the FFT amplitudes of the proton fluxes are multiplied by 100 
and shown. The dashed lines represent 95\% significance levels (blue line for electrons and gray 
line for protons). (c) FFT of the energetic particles near 1.43--1.38 AU (MAVEN: SEP observations) 
where the electron spectra are shown in red and the proton spectra are shown in gray. Note that the 
FFT amplitudes of the proton fluxes are multiplied by 100 and shown. The dashed lines represent 95\% 
significance levels (orange line for electrons and gray line for protons).} \label{fig-fft} 
\end{figure}

\section*{Solar Energetic Particle Observations} \label{sec:SEPObs}

Figure 4a shows the variation of the energetic electron flux in the energy range from $\sim$30 keV to $\sim$315 keV and Figure 4b shows the variation of the energetic proton flux in the energy range from $\sim$320 keV to $\sim$4.8 MeV, both observed by the EPAM sensor onboard ACE satellite located at L1 point near Earth. Four distinct peaks are seen in the electron fluxes upto 100 keV, and all these enhancements were significantly above the quiet background levels. The proton fluxes are smaller in magnitude compared to the electron fluxes during all these major peaks (Figure 4b). However,  it may also be noted that, the proton enhancement signature is also prominent during 3 August (DN 3), 10 October (DN 71) and 16 October 2016 (DN 77), and there is a still  minor proton enhancement during 17 September 2016 (DN 48).\\

Figure 4c shows the variation of the energetic electron flux in the energy range from $\sim$30 keV to $\sim$200 keV and Figure 4d shows the variation of the energetic proton flux in the energy range from $\sim$340 keV to $\sim$5 MeV, both observed by the SEP instrument  onboard MAVEN  located at 1.43--1.38 AU near Mars.  Here also the signature of the periodic enhancements is seen as four distinct peaks in the electron fluxes, in all the three channels upto $\sim$200 keV.  Figure 4d shows the corresponding proton enhancements, measured by MAVEN/SEP. Here also, the proton fluxes are  smaller in magnitude compared to the electron fluxes during all the major peaks, but the enhancements are clearly above the background quiet time fluxes in all channels. Near 1.43--1.38 AU, the fluxes are enhanced by a factor of $\sim$20, compared to the proton fluxes observed at 1 AU. Electrons also show enhancement as they reach 1.43--1.38 AU, but to a lesser extent ($<$10 times, between 1 and 1.43--1.38 AU). Among the four enhancements, the first and third SEP enhancements could have contributions from CME eruptions during the beginning of August and during the end of September. The second and fourth enhancements can be considered as only due to CIR events. However, as mentioned earlier, there are eruptions during the end of September, which are mostly associated with the west\textendash limb CME event and the SEPs arrive at the observers through magnetic connection of the CME accelerated shock source. The magnetically connected SEP intensities are often found to be order of magnitude lower than the SEPs directly coming to the observer location \citep{Xie2017}, and therefore, the contributions of these to the major CIR/SIR related enhancements would be secondary in nature. \\

\begin{sidewaysfigure}[ht]
\centering
\includegraphics[scale=0.5]{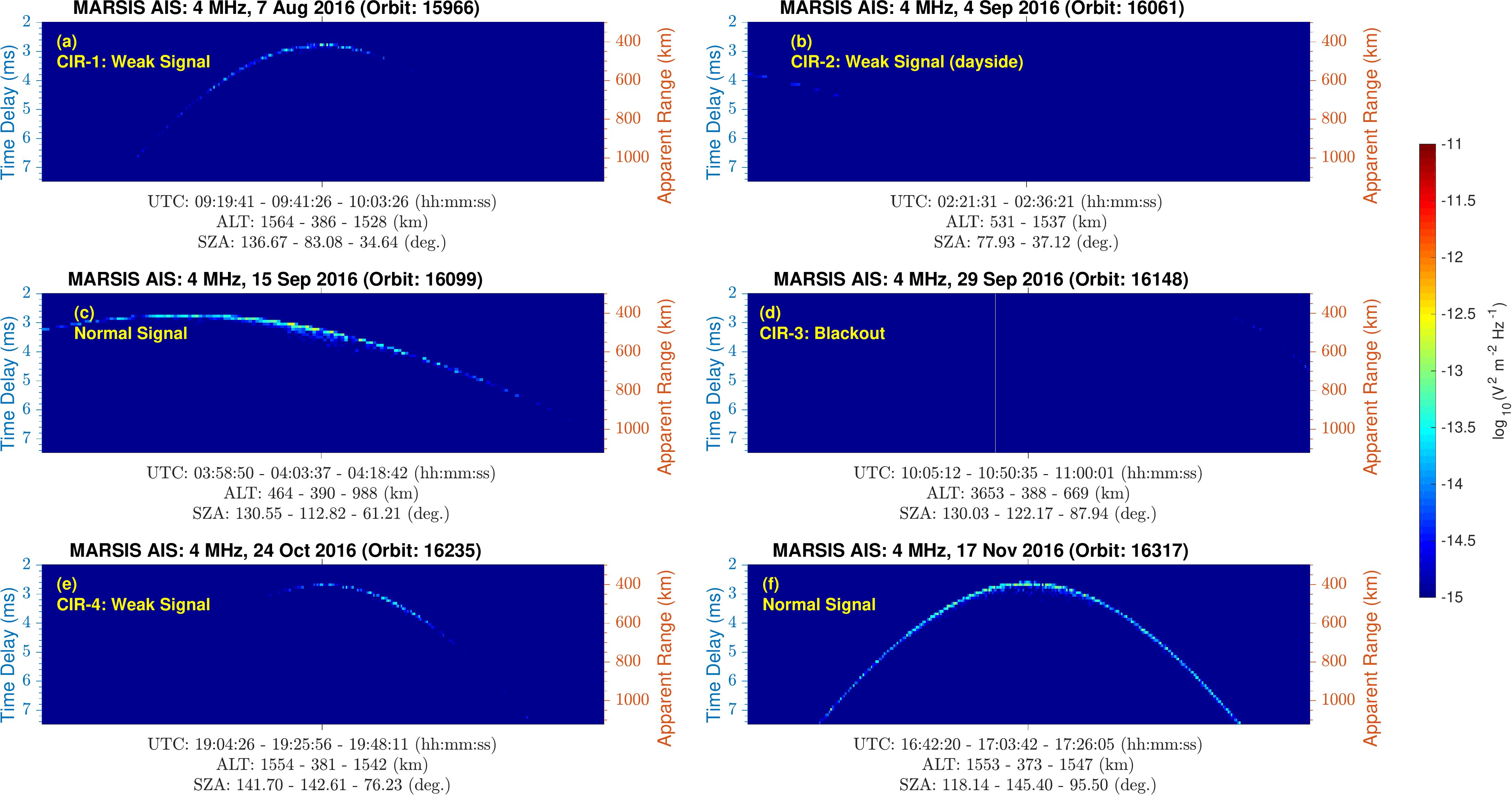} 
\caption{MARSIS Active Ionospheric Sounding (AIS) mode observations. The variation of the signal 
levels at 4 MHz is shown. Representative examples showing normal signals, blackouts, and weak 
signals are given. The time delay experienced by the signal is shown as the left y-axis and the 
apparent range is shown in the right y-axis. The along track of change in satellite altitude and 
Solar Zenith Angle (SZA) are also shown. The color bar represents the signal strength in 
units of electric field spectral density.} \label{fig-marsis} 
\end{sidewaysfigure}

Apart from the four major enhancements related to CIR arrivals mentioned above, there are two other 
enhancements observed. The first one peaks on 18 September (DN 49) and the second enhancement peaks on 15 October (DN 76) at Mars (Table 1). Electron enhancements are also seen during these days, but with much 
shorter duration. These enhancements are possibly related to the interplanetary shock during 
17\textendash20 September (DN 48 to 51) and a weak CME eruption on 8\textendash 9 October, 2016 (DN 
69 to 70, Figure 1e), whose SEPs arrive at Earth and Mars on 12 and 15 October (DN 73 and 76) 
respectively (Table 1). It is interesting to see that a very weak CME could enhance SEPs to a 
significant level. The peak of these enhancements are only approximately an order of magnitude lower 
than the SEP event during March 2015 which is considered to be a major CME event observed at Mars by 
MAVEN \citep{Jakosky2015b, Thampi2019}. Other minor enhancements in solar wind parameters and 
energetic particle flux during the period could be due to the transient fast solar wind and 
accelerated population of interplanetary/solar wind particles \citep{Prinsloo2019}.\\ 



To estimate the periodicities and the amplitudes of the signals with different frequencies 
(periods), we have performed a Fast Fourier Transform (FFT) analysis of the solar wind at L1 point 
and the SEP fluxes observed near 1 and 1.43--1.38 AU. Daily values are used for this analysis after removal 
of the mean.  The solar wind speed observations near Mars were not subjected to this analysis 
because of the limited data length. Figures 5a shows the FFT spectra of the solar wind at L1 point 
near Earth. Quasi\textendash27, quasi\textendash13 and quasi\textendash9 day periodicities are 
significantly observed, as previously reported by \citet{TulasiRam2010} for the CIR observations 
during 2008 period, which was the declining phase of the previous solar minimum. Figure 5b shows 
the FFT spectra of SEP electrons, and protons (for protons with amplitudes multiplied by 100, for 
clarity in visualization) observed by ACE.  The electron flux spectra shows the highest amplitude 
for the quasi\textendash27 day, followed by quasi\textendash13 day period.  Figure 5c shows the FFT 
spectra of SEP electrons and protons (shown in the figure with multiplication by 100 for protons) 
observed by MAVEN. Here also, the electron flux spectra show the highest amplitude for the 
quasi\textendash27 day, followed by quasi\textendash13 day periodicity. This confirms that the SEP 
electron enhancements are predominantly caused by the SIR/CIR during this period, and the transient 
CME eruptions could not distort the periodic nature of the electron enhancements.\\

The periodicities in proton fluxes near 1 AU (Figure 5b) are of much lesser amplitude compared to 
that of electrons, but quasi\textendash27 day is seen to be most prominent. Interestingly, the FFT 
analysis of proton-fluxes near 1.43--1.38 AU (Figure 5c) shows that the quasi\textendash27 day periodicity 
is completely diminished, whereas the quasi\textendash13 day and quasi\textendash9 day periods are 
above the 95\% significance level. Here also, the periodicities are of lesser amplitude compared to 
that in electron fluxes. Here, the weak transient CME eruptions and their interaction with SIRs 
could diminish/completely distort the periodic nature of the proton enhancements.\\

\begin{figure}[h] \hspace{-1cm} \includegraphics[scale=1.5]{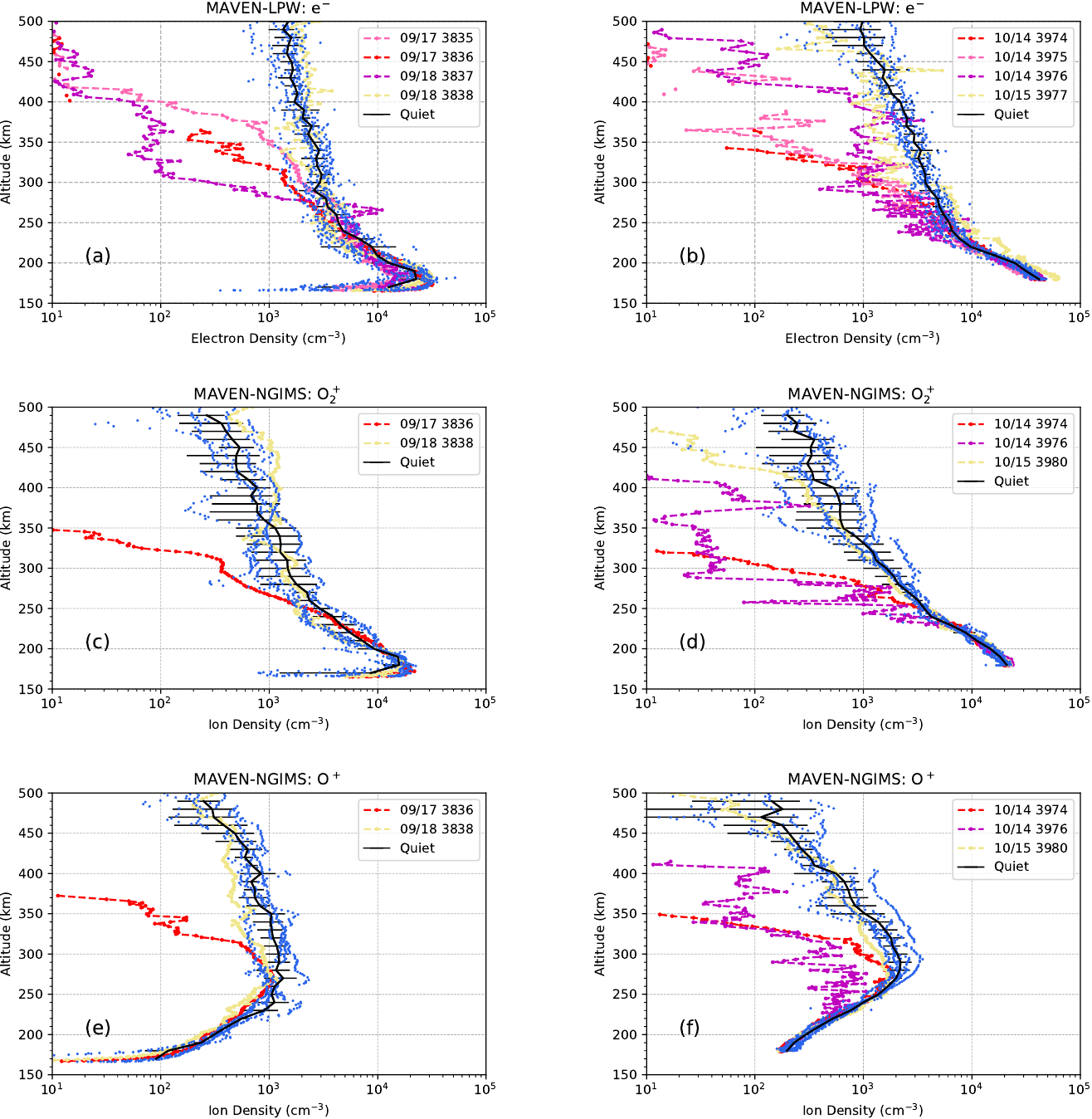} 
\caption{(a), (c), and (e): MAVEN observations for September 2016 CME event: (a) LPW measurements of electron density variation during the CME period - 17 and 18 September 2016 (DN 48 and 49) along with `Quiet' orbits - 15, 16, and 20 September 2016 (DN 46, 47, and 51), their mean and standard deviations. (c) and (e) NGIMS O$_2^+$ and O$^+$ variation during the CME period - 17 and 18 September, 2016 (DN 48 and 49) along with `Quiet' orbits - 15, 16, and 20 September 2016 (DN 46, 47, and 51). (b), (d), and (f): MAVEN observations for October 2016 event: (b) LPW measurements of electron density variation during the CME period - 14 and 15 October, 2016 (DN 75 and 76) along with `Quiet' orbits - 12, 13 and 16 October, 2016 (DN 73, 74, and 77), their mean and standard deviations. (d) and (f) NGIMS O$_2^+$ and O$^+$ variation during the CME period - 14 and 15 October, 2016 (DN 75 and 76) along with `Quiet' orbits - 12, 13 and 16 October, 2016 (DN 73, 74, and 77).} \label{fig-ngims} \end{figure}

\section*{Impacts on Mars} \label{sec:impactsObs}

Figure 6(a-f) show examples of the  variation of 4 MHz signal of the MARSIS radar during the AIS mode operations. The time delay experienced by the signal is shown as the left y-axis and the apparent range is shown in the right y-axis. The along\textendash track change in satellite altitude and Solar Zenith Angle (SZA) are also shown. The color bar represents the signal strength in units of spectral density. Reflected signals at these frequencies ($>$3 MHz) are basically reflections from the Martian surface \citep{Morgan2014}, especially for solar zenith angles $>$70 deg.  Representative examples showing normal signals, blackout, and weak signals are given. During the CIR-3 arrival, there is a complete blackout of the radio signal, whereas during the CIR arrivals 1, 2, and 4, the signals were significantly weak, indicating partial absorption. However, observations during CIR-2 were on the dayside and close to terminator, where signal attenuation at SZA$<$70 deg. could be mostly due to dayside absorption, which is true for all observations at lower solar zenith angles. The variation of signals at other frequencies above 3 MHz such as 3.5 MHz, 4.5 MHz, and 5 MHz are similar to 4 MHz (hence not illustrated).\\

Figure 7  shows the impact of the interplanetary shock and weak CME events on the Martian ionosphere. The left panels (Figure 7(a, c, e)) shows the LPW -- e$^-$, NGIMS -- O$_2^+$ and O$^+$ profiles corresponding to the September 17-18 (DN 48-49) interplanetary shock event, and the right panels (Figure 7(b, d, f)) shows the LPW -- e$^-$, NGIMS -- O$_2^+$ and O$^+$ profiles corresponding to the  October 14-15 (DN 75-76) stealth CME event (caused by the CME eruption during 8-9 October).  Figure 7a shows  the electron density profiles during 17 to 18 September 2016, along with the  quiet time profiles (several orbits on 15, 16 and 20 September) with their mean and standard deviation. Orbits 3835, 3836 (September 17) and 3837 (September 18) show deviation compared to the mean quiet time profile. The ionopause (the ionospheric boundary, where the ionospheric pressure balances the solar wind dynamic pressure) altitude observed for these three orbits are below 400 km. The next orbit, 3838 on September 18 shows a normal behavior. The O$_2^+$ and O$^+$ profiles (Figures 8c and 8e) also show a similar variation, with ionopause below 400 km. It must be noted that NGIMS alternates between ion and neutral modes, whereas LPW measures the electron density in all orbits, and hence the signature is seen only in one orbit in NGIMS data. The magnetic field started to enhance from the quiet time values on 17 September $\sim$18 UTC, and maximized at 23 hour 40 minutes UTC to 13.3 nT. The response seen in LPW profiles are concurrent to these enhancements. Similar features can be seen during the CME event (14-15 October 2016) as well. Here the typical quiet orbits are taken from several orbits on October 12, 13, and 16. The LPW electron data during the event orbits show deviation from the quiet time profiles for three consecutive orbits (Figure 7b) and the NGIMS O$_2^+$ and O$^+$ profiles show differences from quiet time behavior in two profiles (Figures 7d and 7f), on 14 and 15 October. Though solar wind magnetic field data are not available during this period, from the arrival times of the CME and the analysis on the CME-HSS interaction reported by \citet{He2018}, it can be concluded that the depletion and compression seen in the topside ionosphere are the consequence of the weak CME. The shock of the CME structure arrives first along with the energetic particles, and subsequently the CME sheath region arrives with fluctuations in the IMF. The compression of the Martian magnetosphere--ionosphere system is observed during the entire period of passage of CME structure and associated energetic particles \citep{Jakosky2015b}.\\

\section*{Discussion}

The CIRs are often associated with shocks, when electrons are accelerated, as observed by the Ulysses spacecraft when it was exploring the three-dimensional heliosphere between 1 and 5 AU during the declining phase of solar activity \citep{Mann2002}. A field\textendash aligned flux of energetic electrons with energies up to $\sim$ 200 eV or higher is commonly observed upstream from  both the forward and the reverse shocks that bound CIRs at heliocentric distances greater than $\sim$2 AU \citep{Gosling1993}. The Ulysses  spacecraft observations confirmed that majority of the periodic energetic particle increases are due to particles accelerated at long\textendash  lived CIRs   which persisted over several solar rotations. During all the four enhancements observed in the present study, electron fluxes measured near 1 AU and 1.43--1.38 AU are higher compared to that of ions, indicating that these are electrons which are accelerated at the reverse shock and re-entered the inner heliosphere which mirrors to the reverse shock and get repeatedly  accelerated. This process is less effective for the ions \citep{Simnett1995} which may explain the much lower ion fluxes. These observations provide  evidence that the SEP electrons that are accelerated at the reverse shock mainly associated with CIRs, which counter stream to the inner heliosphere and again mirrors to the reverse shock to get significantly accelerated, retains their recurrent nature even in the presence of intermittent slow CMEs whereas the observed energetic proton enhancements are probably more influenced by the CME-SIR interactions.\\ 

The SEP electrons caused blackout/weakening of radio signals, as evident from MARSIS observations. Previously long duration HF radar blackout has been observed  by MARSIS and SHARAD (Mars SHAllow RADar sounder) during the very intense space weather event in September 2017 \citep{Sanchez2019}. The loss of signal was interpreted to be  due to the formation of an ionospheric layer near $\sim$90 km, that absorbed HF radar signals similar to  the D\textendash region absorption in the terrestrial ionosphere. Theoretical calculations showed that such an ionospheric layer is created by SEP electrons, rather than previously proposed SEP protons \citep{Sanchez2019}. Though the recurrent geomagnetic activity due to CIRs were previously reported for Earth's ionosphere \citep{Tsurutani1995, TulasiRam2010}, this kind of a recurrent occurrence of energetic electron enhancement associated with CIRs and their impacts are so far not reported from a vantage point near Mars, except the report by \citet{Morgan2010}, where they showed two events of the absorption of radar signals, separated by an interval of approximately 27 days, corresponding to one solar rotation. In this case, the directly detected ion enhancements were only observed during the first event, whereas energetic electron enhancements were not reported. The present events clearly indicate that periodic radar blackouts during CIRs are associated solar energetic electron enhancements and this could be a major space weather impact on Mars during the declining phase of the solar activity.\\

The interplanetary shock associated with the HSS during September 17-18 (DN 48-49) could significantly alter the Martian plasma environment by compressing the ionospheric boundary to lower altitudes (Figure 7a,7c,7e). The geo-effectiveness of  this interplanetary shock at Earth was less primarily due to the magnetic field configuration, because the geomagnetic activity is linked to the southward IMF \citep{Gosling1991}. This period was also characterized by a HSS-stealth CME interaction event (October 12-15). The CME-HSS interaction during this event has been studied in detail by \citet{He2018}, which shows a weak CME without obvious signatures in the low corona could produce a relatively intense geomagnetic storm at Earth. Within the ICME, the total B was as high as 25 nT while the southward IMF reached -21 nT, on 13 October \citep{He2018}. During this event, the CME propagating into an ambient solar wind is followed by the HSS ($\sim$500 km/s) originating from the equatorial coronal hole. Previously, it has been reported that the presence of a preceding CME a few hours before a fast eruption has been found to be connected with higher fluxes of solar energetic protons, whereas the CME-CME interaction occurring within the solar corona is often associated with unusual radio bursts, indicating electron acceleration \citep{Gopalswamy2002, Gopalswamy2004, Lugaz2017b}. It has also been reported that the trajectory of CMEs is significantly affected when the eruptions occur in close proximity to coronal holes, and lead to driver-less shocks due to purely geometrical reasons \citep{Gopalswamy2009}. In the present case, the interaction of the weak CME with SIR/CIR could be the cause of the enhanced SEP fluxes associated with the weak CME events. It may be noted that \citet{He2018} provides evidences for the CME\textendash SIR/CIR interaction for the October event, and show that the very slow CME of 8\textendash 9 October, 2016 is compressed since it is bracketed between a slow and fast wind, and hence produce a geomagnetic storm at Earth. Our study shows that the Martian topside ionosphere was also impacted by this event (Figure 7b,7d,7f).\\

The observations that the Martian ionopause-like density gradient is seen at lower altitudes during CME events \citep{Vogt2015,Thampi2018} as well as during CIRs \citep{Lee2017, Krishnaprasad2019} is well understood. However, all these previous reports were for intense CME and CIR periods (like March 2015, June 2015, and September 2017 event periods). \citet{Sanchez2017} studied the response of Martian plasma environment to an ICME during a solar minimum period in March 2008, whose dynamic pressure was less than 10 nPa, based on proxy measurement from 1 AU by STEREO-B. A large compression of magnetosphere and ionosphere was observed but for a smaller scale compared to events during moderate and high solar activity conditions.  The unexpectedly new observation here is that topside ionospheric response to a similar extent to these intense events is caused by a relatively weak CME and an interplanetary shock. The maximum dynamic pressure for the interplanetary shock event was $\sim$9 nPa, while the peak dynamic pressure of 8 March 2015 CME event was $\sim$15 nPa. However, as mentioned earlier, we do not have upstream dynamic pressure observations during the stealth CME event. This shows that the interplanetary evolution of CMEs including slow ones can be quite complicated, such as interactions with the highly structured solar wind \citep{Gopalswamy2001, Lugaz2012}. The present study shows that as the combination of different solar wind structures can result in enhanced geo-effectiveness \citep{Liu2014}, the impact on non-magnetized planets like Mars can also be enhanced by the CME-HSS interaction. This will have significant implications in the calculations of ion-escape from Mars since weak or declining phase of solar cycle does not necessarily suggest low impact or less erosion at the topside Martian ionosphere.

\section*{Summary} \label{sec:Summ}

Recurrent energetic particle enhancements are observed at Earth and Mars, both were nearly radially aligned in the ecliptic plane, during the period between 1 August and 15 November 2016 in the declining phase of solar cycle 24. These corotating events are high speed streams forming SIR/CIR, and are due to the appearance of a stream producing coronal hole in the solar disk for four consecutive solar rotations. Several spacecraft such as ACE and MAVEN had simultaneously observed the arrival of SEPs during this period, with major periodicities of quasi $\sim$27 day and quasi $\sim$13 day. These CIR events were prominently energetic electron events, probably due to the reverse shock acceleration of electrons in CIR. Radio blackout/partial absorption of radio signals at Mars were observed by MARSIS radar onboard MEX during the energetic electron enhancements. These observations suggests that solar energetic electrons could be causing radio blackouts in contrast to proton rich events, as reported by \citet{Sanchez2019} using numerical calculations.\\

Interestingly, an HSS-related interplanetary shock and a weak CME, in September and October 2016 respectively, had interacted with the SIR/CIR structures forming merged interaction region in the inner heliosphere. These interaction events produced enhancement in proton intensities at Earth and Mars, making the weaker CME and the interplanetary shock to produce larger impacts at these planets, such as decreased ionopause plasma boundary height at Mars as observed by LPW and NGIMS onboard MAVEN. The interaction also caused quasi\textendash27 day periodicity to be completely diminished for solar energetic protons at 1.43--1.38 AU (near Mars) in comparison to 1 AU (near Earth).

\acknowledgments
The work is supported by the Indian Space Research Organisation (ISRO). The ACE EPAM/SWEPAM/MAG data are  taken from the ACE Science Center (\url{http://www.srl.caltech.edu/ACE/ASC/}). We thank the staff of the ACE Science Center for providing the ACE SEP, IMF, and solar wind data. The MAVEN data used in this work are taken from the NASA Planetary Data System (\url{https://pds.nasa.gov/}). We gratefully acknowledge the MAVEN team for the data. The MEX/MARSIS data are taken from the ESA Planetary Science Archive (\url{https://www.cosmos.esa.int/web/psa/mars-express}). We acknowledge the MEX/MARSIS Principal Investigators, G. Picardi (Universita di Roma \lq La Sapienza\rq, Italy), R. Orosei (IAPS, Rome, Italy), and J. Plaut (JPL, Pasadena, USA) as well as the ESA Planetary Science Archive (\url{https://archives.esac.esa.int/psa/}) for the data. We also acknowledge using solar observations from SOHO/LASCO (\url {https://cdaw.gsfc.nasa.gov/CME_list/}) and SDO/AIA (\url{https://sdo.gsfc.nasa.gov/}). The WSA-ENLIL model simulations are used from \url{https://helioweather.net/}. C. Krishnaprasad acknowledges the financial assistance provided by ISRO through a research fellowship. This research has made use of SunPy v2.0, an open-source and free community-developed solar data analysis Python package (\url{https://sunpy.org/}).



\end{document}